\documentclass{aastex701}

\begin{document}

\title{Identification of a Large-Scale Diffuse Gamma-Ray Structure in the Southern Galactic Hemisphere}
\author[orcid=0009-0005-6541-214X,sname='Xie']{Zhen Xie}
\affiliation{School of Astronomy and Space Science, University of Science and Technology of China, Hefei, 230026, China}
\email[]{}
\author[orcid=0000-0002-2750-3383,sname='Huang']{Xiaoyuan Huang}
\affiliation{Key Laboratory of Dark Matter and Space Astronomy, Purple Mountain Observatory, Chinese Academy of Sciences, Nanjing 210023, China}
\affiliation{School of Astronomy and Space Science, University of Science and Technology of China, Hefei, 230026, China}
\email[show]{xyhuang@pmo.ac.cn}
\author[orcid=,sname='Liu']{Bing Liu}
\affiliation{Key Laboratory of Dark Matter and Space Astronomy, Purple Mountain Observatory, Chinese Academy of Sciences, Nanjing 210023, China}
\affiliation{School of Astronomy and Space Science, University of Science and Technology of China, Hefei, 230026, China}
\email[]{}
\author[orcid=0000-0001-5801-2547,sname='Yang']{Ruizhi Yang}
\affiliation{School of Astronomy and Space Science, University of Science and Technology of China, Hefei, 230026, China}
\email[show]{yangrz@ustc.edu.cn}

\correspondingauthor{Xiaoyuan Huang, Ruizhi Yang}

\begin{abstract}
We identify and characterize a large-scale diffuse gamma-ray structure in the Southern Galactic Hemisphere using 17 yr of Fermi-LAT data. An energy-dependent likelihood analysis, including alternative Galactic diffuse-emission models, isotropic emission, the Fermi bubbles, and resolved 4FGL sources, reveals an extended excess that persists across the tested background models and spans tens of degrees. The excess broadly follows the X-ray-defined southern eROSITA Bubble (eB) region, while also overlapping the projected southern extension of Loop I. Template fits favor a filled eB-like morphology over the adopted Wolleben Loop I shell geometry, making the structure a plausible gamma-ray counterpart of the southern eB, although Loop-I-related or other localized foreground emission cannot be excluded. Under the eB template, the southern component is fainter and softer than the northern large-scale component, with an integrated luminosity lower by a factor of about seven, broadly consistent with the eROSITA-bubble asymmetry. If interpreted as Galactic-scale outflow emission, its faint, soft spectrum may indicate aged particles and/or distributed reacceleration in the outer bubble. A hadronic interpretation is energetically demanding, whereas a leptonic inverse-Compton scenario is more economical but requires rapid transport and/or local reacceleration of high-energy electrons.

\end{abstract}


\section{Introduction}
The origin and physical nature of the giant, high-latitude structures flanking the Galactic Center have remained one of the most intriguing puzzles in high-energy astrophysics. Early multi-wavelength observations, ranging from mid-infrared to microwave and X-ray bands, first hinted at large-scale Galactic outbursts \citep{bland2003large, finkbeiner2004microwave, akita2018diffuse}.
The discovery of the Fermi bubbles (FBs), two giant gamma-ray emitting lobes extending $\sim 10$ kpc symmetrically with respect to the Galactic Center (GC), has revolutionized our understanding of energetic feedback in the Milky Way \citep{su2010giant,crocker2011fermi,guo2012fermi,yang2014fermi,ackermann2014spectrum}. Characterized by a hard spectrum and uniform surface brightness, the FBs are widely interpreted as relics of past nuclear outbursts, driven either by intense starburst activity or episodic accretion onto Sagittarius A* \citep{yang2022fermi, sarkar2024fermi}. Beyond these canonical bubbles, the high-latitude gamma-ray sky reveals a complex landscape of diffuse structures, many of which appear to correlate with well-known radio and X-ray features, such as the North Polar Spur (NPS) \citep{haslam1982408,sofue1979radio,snowden1997rosat}, WMAP-Planck haze \citep{dobler2012last,bartlett2013planck} and the broader Loop I complex \citep{haslam1982408, kataoka2018x,Mertsch:2013pua,Liu:2014mpa,vonHausegger:2015flg}. 

The multi-wavelength perspective has further expanded this picture. The eROSITA all-sky survey recently uncovered the eROSITA bubbles (eBs), large X-ray structures that encompass the FBs and extend up to 14 kpc into the Galactic halo \citep{predehl2020detection}. While the FBs and eBs exhibit a striking morphological alignment, suggesting a unified Galactic outflow, the nature of their larger-scale environment remains a subject of intense debate. Historically, features like the NPS were attributed to local supernova remnants within the solar neighborhood ($\sim 100$ pc), as proposed by the ``two-shell" model \citep{wolleben2007new}. However, the apparent symmetry of the eBs across the Galactic plane and observed X-ray shadowing effects have strengthened a Galactic-scale interpretation, where these structures represent limb-brightened edges of a massive biconical outflow originating from the GC \citep{sarkar2024fermi, Liu2024}.

A critical missing piece in this Galactic-scale framework is the
apparent north--south asymmetry in the gamma-ray regime. While a clear
gamma-ray counterpart to the NPS has been identified in the northern
sky \citep{ackermann2014spectrum}, no comparable southern
large-scale component has been firmly established and systematically
characterized. \citet{sarkar2019possible} predicted faint southern
X-ray and gamma-ray emission, while
\citet{scheel2023multicomponent} showed extended gamma-ray emission in
the southern sky that may be part of a more continuous structure
surrounding the Fermi bubbles. These results provided earlier
theoretical and observational indications of a southern component, but
its morphology and spectrum were not quantified through a dedicated
energy-dependent template likelihood analysis across alternative
Galactic diffuse-emission models. This apparent asymmetry poses a significant challenge. If the
eBs trace a Galactic-scale bipolar outflow, one may expect a
corresponding population of shock-accelerated cosmic rays (CRs) emitting
gamma-rays in the south. It remains unclear whether the lack of
a clearly established southern counterpart reflects intrinsic
physical asymmetries in the circumgalactic medium or whether
the southern emission was simply too faint to be distinguished from
the intense Galactic diffuse background in previous analyses.

In this paper, we leverage 17 years of Fermi Large Area Telescope (Fermi-LAT) Pass 8 data to perform a sensitive search for diffuse gamma-ray structures in the sky. By disentangling various diffuse components through energy-resolved template fitting and testing alternative Galactic diffuse-emission models, we identify and characterize a broad southern Galactic-hemisphere emission structure that spatially overlaps the southern eROSITA Bubble region. Its morphology and spectrum provide new clues to the possible high-energy counterpart of the southern eROSITA Bubble and to the role of Galactic-center-driven outflows in shaping the high-latitude gamma-ray sky.

The paper is organized as follows. In Section \ref{sec2}, we detail our data analysis pipeline and methodology. This includes the selection of 17 years of Fermi-LAT Pass 8 data, the optimization of background models, with particular attention to the use of alternative Galactic diffuse emission models, and the component separation techniques employed to isolate faint, large-scale structures. Section \ref{sec3} presents our primary observational results, focusing on the morphological identification of the detected southern structure. Its spatial association with known features and the resulting physical implications are discussed in Section \ref{sec4}, which also contains our summary and concluding remarks. Detailed descriptions of the background modeling and spectral analysis are provided in Appendices \ref{ap:bg} and \ref{apsed}.

\section{Data Analysis}\label{sec2}
\subsection{Fermi-LAT data}
In this study, we utilize approximately 17 years of Fermi-LAT data, covering the period from August 4, 2008, to October 20, 2025 (mission elapsed time (MET) 239,557,417 to 782,614,950). 
To ensure a high-purity photon sample suitable for the analysis of diffuse large-scale structures while maintaining sufficient statistics, we select events from the \texttt{CLEAN} event class (\texttt{evclass=256}) within the energy range of 200\,MeV to 500\,GeV. 
To guarantee data quality, we apply standard selection criteria by keeping time intervals when the spacecraft was in a valid science mission configuration and the data were flagged as good (\texttt{DATA\_QUAL > 0 \&\& LAT\_CONFIG == 1}). Furthermore, to mitigate contamination from the bright Earth-limb emission, we exclude photons with zenith angles greater than $90^\circ$. Finally, the exposure correction is performed following the standard Fermi-LAT analysis procedures to obtain the finalized all-sky counts map.

\subsection{Method of analysis}

The remaining high-latitude data are spatially binned using the HEALPix \citep{2005ApJ...622..759G} scheme with a resolution of $N_{\rm side} = 256$ and divided into 25 logarithmically spaced energy bins from 200\,MeV to 500\,GeV. This configuration corresponds to a total of 786,432 pixels with a mean spacing of approximately $0.23^\circ$ per pixel. Such a high-statistics, multi-year dataset provides the requisite sensitivity to search for faint, extended gamma-ray features across the sky.

The diffuse Galactic emission (DGE), which includes neutral pion
(\(\pi^0\)) decay, bremsstrahlung, and inverse-Compton scattering, is
modeled using a systematic suite of templates generated by the GALPROP
code \citep{Porter_2022}. To account for uncertainties in CR
propagation and the interstellar medium, we consider models spanning
different CR source distributions, halo heights, and interstellar gas
distributions. The retained combinations of these model choices define
an ensemble of 64 DGE templates, each of which is fitted independently
to the LAT data. A representative benchmark model is adopted for the
main analysis, with comparisons to alternative models provided in
Appendix~\ref{ap:bg}.

In addition to the DGE, our model includes templates for the FBs \citep{su2010giant}, the isotropic diffuse background, and a component representing resolved point sources based on the Fermi-LAT 14-year catalog (4FGL; \cite{Abdollahi_2022,ballet2024fermilargeareatelescope}). The 4FGL source component is included as a template and fitted simultaneously with the other emission components in each energy bin. All templates are convolved with the energy-dependent point spread function (PSF) of the Fermi-LAT instrument and projected onto the HEALPix grid with $N_{\rm side}=256$.

To mitigate potential contamination from the brightest, variable, or imperfectly modeled sources, we further apply a spatial mask to the dataset. Specifically, we exclude the Galactic plane region ($|b| < 15^\circ$), as well as circular regions of $2^\circ$ radius around the 370 brightest 4FGL sources (with $\mathrm{TS} > 2500$). The emission regions of known extended sources are also masked using their official spatial templates. These masks are applied consistently to both the data and the model templates.

For each energy bin, we perform an independent maximum likelihood fit to determine the best-fit normalization of each component. The likelihood is constructed assuming Poisson statistics for the photon counts in each spatial pixel. The DGE, isotropic emission, FB template, and 4FGL source template are fitted simultaneously, allowing for a robust decomposition of the observed emission. This DGE + isotropic + FB + 4FGL source fit defines the baseline model used for the residual maps and for the subsequent large-scale template tests.

\section{Results}\label{sec3}
\begin{figure}[htbp]
    \centering
    \includegraphics[width=1\linewidth]{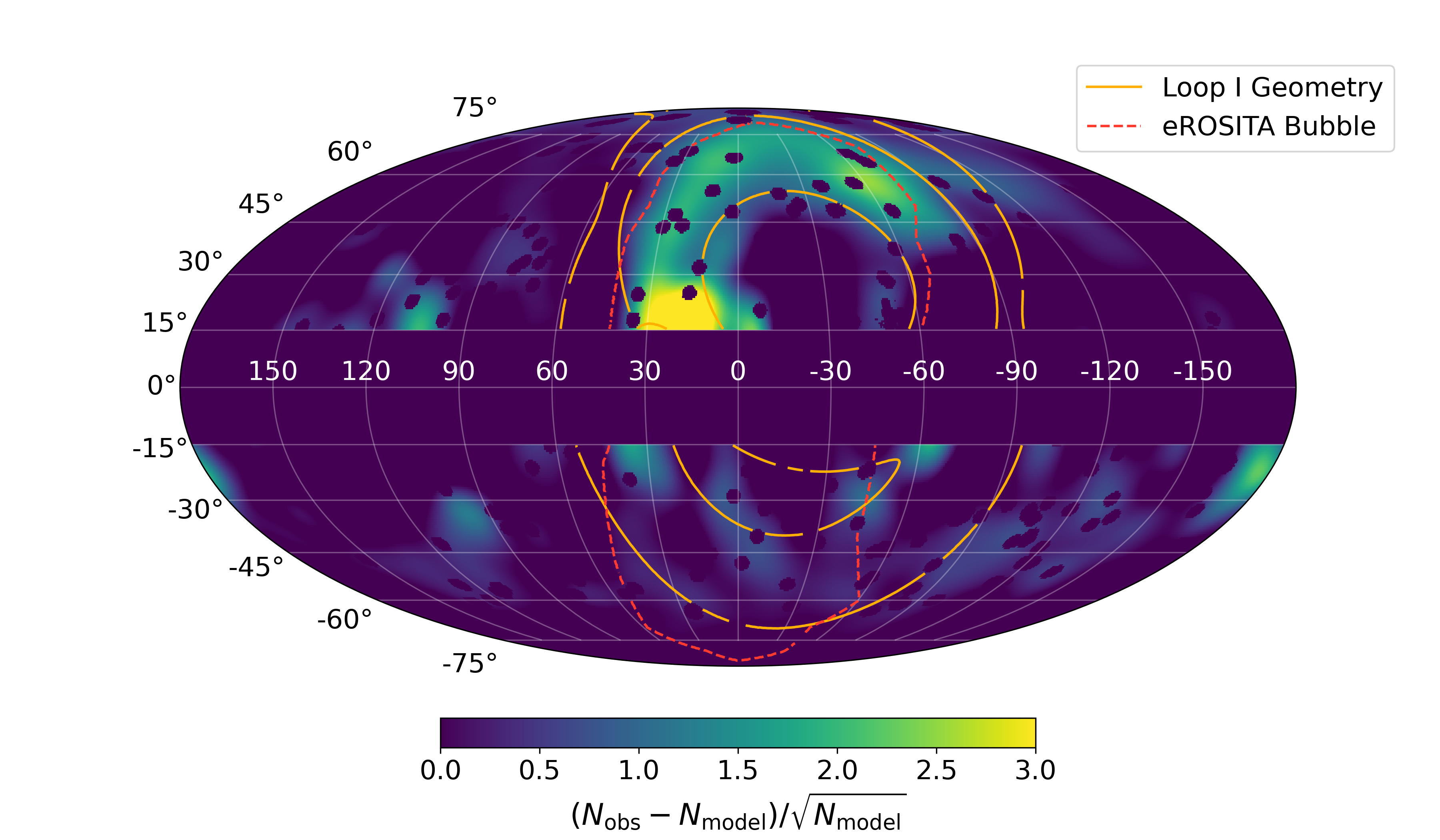} 
\caption{
{All-sky approximate residual-significance map of the large-scale gamma-ray residuals. This map is intended only to visualize the large-scale residual morphology after baseline-template subtraction and is not used as a formal detection statistic. The background color shows $(N_{\rm obs}-N_{\rm model})/\sqrt{N_{\rm model}}$, constructed from the residual counts after the template fit and summed over the analyzed energy bins. The northern and southern hemispheres are fitted separately and merged for visualization. A $10^\circ$ Gaussian smoothing kernel has been applied to highlight extended features.}
The overlaid contours represent the spatial templates for Loop I (yellow solid lines, \cite{wolleben2007new}) and the eBs (red dashed lines, \cite{predehl2020detection}). The Galactic plane region ($|b| < 15^\circ$) is masked to reduce contamination from foreground modeling uncertainties, with all masked regions assigned a numerical value of zero for visualization purposes.}
    \label{fig:skres}
\end{figure}

{After subtracting the fitted baseline templates, we construct an approximate residual-significance map, defined as 
$(N_{\rm obs}-N_{\rm model})/\sqrt{N_{\rm model}}$, where $N_{\rm obs}$ and $N_{\rm model}$ are the observed and best-fit model counts, respectively, summed over the analyzed energy bins. The northern and southern hemispheres are fitted separately and then merged for visualization. A Gaussian smoothing kernel with $\sigma=10^\circ$ is applied to the final map to highlight faint, extended residual features, as shown in Figure~\ref{fig:skres}. This map is used primarily to display the large-scale residual morphology, and the quantitative significance of the large-scale components is evaluated below through template likelihood fits.}

In the northern hemisphere, our analysis clearly recovers the well-known gamma-ray counterpart to the NPS and the broader Loop I complex. The successful recovery of these established features serves as a critical benchmark, validating the sensitivity of our multi-template fitting and the robustness of our background subtraction approach.

In the southern hemisphere, a coherent and expansive emission structure is identified, extending from the Galactic mid-latitudes toward the South Galactic Pole. This feature appears broadly spatially consistent with both the southern lobe of the eBs and the projected southern extension of the Loop I structure. Morphologically, the structure manifests as a large-scale residual that exhibits a notable symmetry relative to the prominent diffuse features in the Northern Galactic Hemisphere. Notably, the gamma-ray residual also appears to show substructures broadly consistent with features seen in the eROSITA map, such as the gap near $l \approx 30^\circ$ \citep{predehl2020detection,yeung2026srgerositadiffusesoftxray}.
The gamma-ray boundaries of this southern structure exhibit a broad morphological correspondence to the X-ray edges of the eROSITA Bubbles. This multi-wavelength spatial correlation suggests a potential physical association with broader Galactic outflows, motivating an eB-associated interpretation that is analogous to the large-scale structures observed in the Northern Galactic Hemisphere.
{Given that Loop I and related radio-loop structures are known to produce prominent large-scale foreground imprints at high Galactic latitudes, from radio synchrotron to microwave bands \citep{Mertsch:2013pua,Liu:2014mpa,vonHausegger:2015flg}, we quantitatively evaluate whether the southern excess is better described by a Loop-I-like shell geometry or by an eB-like filled morphology. We therefore introduce two alternative large-scale templates: the Wolleben Loop I geometry \citep{wolleben2007new}, characterized by two interweaving synchrotron-emitting shells, and the eB spatial model, represented by a uniform-intensity filled region bounded by the X-ray contours \citep{predehl2020detection}. The test statistic is defined relative to the baseline model as ${\rm TS}=2(\ln\mathcal{L}_{\rm template}-\ln\mathcal{L}_{\rm baseline})$, with a larger TS indicating a greater improvement over the baseline. Using our benchmark background model, we obtain ${\rm TS}=881.46$ for the Wolleben Loop I template and ${\rm TS}=959.93$ for the eB template, both relative to the same baseline model. The eB and Loop I templates are non-nested spatial hypotheses with the same number of normalization parameters, so their direct comparison should be interpreted as an empirical likelihood preference rather than a simple likelihood-ratio significance. We compare the competing models using ${\rm AIC}=2k-2\ln\mathcal{L}$, where $k$ is the number of free parameters and a lower AIC indicates the preferred model. Equivalently, the benchmark Akaike information criterion (AIC) difference is ${\rm AIC}_{\rm eB}-{\rm AIC}_{\rm Loop~I}=-78.47$, favoring the eB-like morphology.} This preference for the eB template remains consistent across the majority of alternative DGE models tested
(see Appendix \ref{ap:bg} and Fig. \ref{fig:nscp}); over the full 64-model ensemble, the mean template TS values
are $\langle {\rm TS}_{\rm Loop~I}\rangle = 763.0$ and $\langle {\rm TS}_{\rm eB}\rangle = 932.2$. These large template TS values indicate that the fit strongly favors an additional large-scale component in the southern high-latitude sky. Furthermore, 
{the preference for the filled eB-like template over the adopted Wolleben Loop I shell geometry supports an eB-associated interpretation as a plausible scenario, while foreground contributions and large-scale template degeneracies cannot be fully excluded. The filled template is used here as an empirical description of the projected large-scale morphology. In a shock-related interpretation discussed below, emission from a kpc-scale downstream shocked region associated with the eB forward shock, together with post-shock advection, line-of-sight projection, and LAT smoothing, can appear less limb-brightened \citep{predehl2020detection}. In this interpretation, the southern excess could be naturally understood as part of a Galactic-scale outflow.
}

\begin{figure}[htbp]
    \centering
    \includegraphics[width=0.6\linewidth]{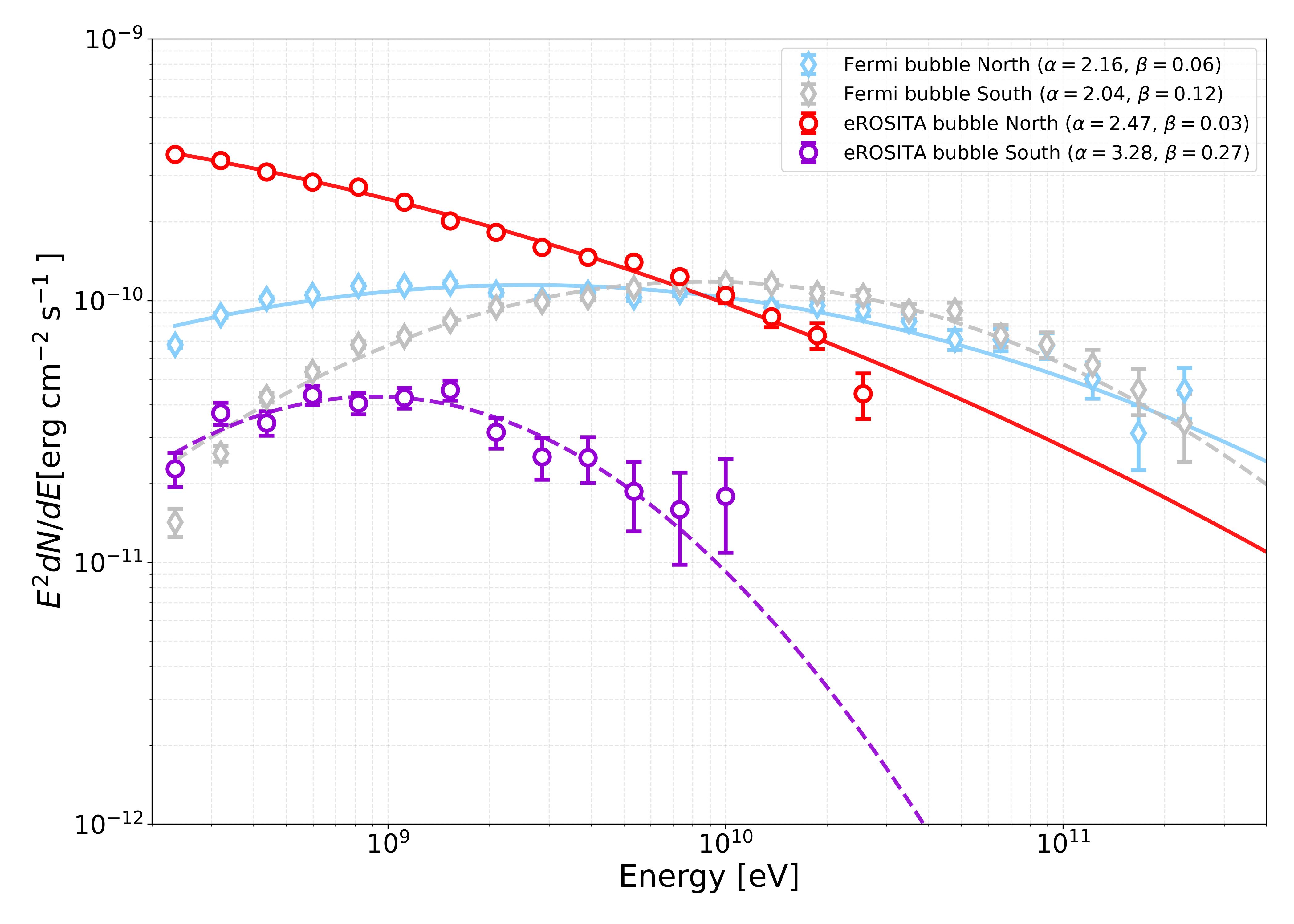}
    \caption{Comparison of gamma-ray SEDs for the northern (solid lines) and southern (dashed lines) large-scale structures in our analysis. Data points represent extracted fluxes for the FBs (diamonds) and eBs (circles). The curves show the best-fit log-parabola models for the corresponding components, and the quoted $\alpha$ and $\beta$ values in the legend denote the best-fit parameters. Error bars represent $1\sigma$ nominal uncertainties.}
    \label{fig:all-spec}
\end{figure}

We extracted the energy spectra for several distinct components, including the FBs and the large-scale structures modeled using eB and Loop I templates, as shown in Fig. \ref{fig:all-spec}. To ensure spectral robustness, data points for the eB and Loop I components are retained only for energy bins with template-fit significance $\sqrt{\mathrm{TS}}>3$. To further quantify the spectral shape, we performed fits using a log-parabola model, defined as $E^2 dN/dE = N_0 (E/E_{\text{pivot}})^{2-\alpha-\beta \ln(E/E_{\text{pivot}})}$, where the pivot energy $E_{\text{pivot}}$ is fixed at 10 GeV.

Based on the best-fit log-parabola model, we integrate the Spectral Energy Distribution (SED) over the energy range of 0.2–500 GeV to derive the gamma-ray flux and the corresponding luminosity under the eB model assumption (see details in Appendix~\ref{apsed}). The total luminosity ratio between the northern and southern structures is approximately 7.2, a value that reflects a pronounced north–south asymmetry. This is broadly consistent with the intensity distribution of the eROSITA bubbles (eBs), which also exhibit a significantly stronger northern component (estimated at approximately 3–6 times the southern intensity based on the luminosity profile integrated over the $\pm 50^\circ$ latitude range \citep{mou2023asymmetric}, with the variation in this ratio primarily arising from differences in background subtraction).

While the spectra of the FBs remain consistent between the northern and southern hemispheres, our results reveal that the larger-scale components exhibit noticeable hemispheric asymmetries. Notably, the southern structure exhibits a softer spectrum compared to its northern counterpart. Furthermore, the emission intensity in the northern hemisphere appears markedly stronger than the diffuse features in the south. This spectral softening and the lower intensity in the south may suggest a more aged cosmic-ray electron population and/or less efficient in situ reacceleration in the southern large-scale component.

To connect the measured gamma-ray spectra with the underlying
particle populations, we perform spectral fits with \texttt{naima}
under both local and Galactic Center distance assumptions, as described
in Appendix~\ref{apsed}. For the adopted BPL electron model, the fitted
high-energy indices are \(\gamma_2=3.53\) and \(3.55\) for the local
and Galactic Center Loop I cases, respectively, and \(4.75\) and
\(4.45\) for the corresponding eB cases. The eB-template SED therefore
requires a softer high-energy electron spectrum. The full leptonic and
hadronic fit parameters are summarized in
Table~\ref{tab:combined_fit}.

\begin{table}[]
  \centering
  \caption{Best-fit parameters derived from the SED analysis of the southern gamma-ray emission. The upper part assumes a leptonic origin (BPL electron distribution), and the lower part assumes a hadronic origin (PL/ECPL proton distribution with $n_H = 1.0\,\mathrm{cm}^{-3}$ for the local case and $10^{-3}\,\mathrm{cm}^{-3}$ for the GC case). Two distance scenarios are considered: local ($d = 400$~pc) and Galactic Center (GC, $d = 8$~kpc). Errors represent $1\sigma$ uncertainties. The reported $\ln L$ values are computed using only the detected data points and do not include contributions from the upper limits.}
  \label{tab:combined_fit}
    \setlength{\tabcolsep}{9pt}
  \begin{tabular}{llcccc}
    \tableline\tableline
    Model & Scenario & Index  & electron break energy / proton cutoff energy & $W_{e/p} (>1 \text{ GeV})$ (erg) & $\max \ln \mathcal{L}$ \\
          &          &       ($\gamma_2 / \alpha$)        & $E_{b,e}$ (GeV)/ $E_{c,p}$ (TeV) &  &  \\
    \tableline
    \multicolumn{6}{c}{Leptonic Origin (BPL)} \\
    \tableline
    Loop I & Local & $3.53^{+0.25}_{-0.17}$ & $6.5^{+2.4}_{-1.8}$ & $5.9^{+2.0}_{-0.8} \times 10^{48}$ & $-4.89$ \\
           & GC    & $3.55^{+0.27}_{-0.13}$ & $(1.8^{+1.0}_{-1.7}) \times 10^{2}$ & $2.8^{+3.7}_{-2.7} \times 10^{52}$ & $-5.72$ \\
    eB     & Local & $4.75^{+0.74}_{-0.79}$ & $14.3^{+1.7}_{-1.7}$ & $2.20^{+0.21}_{-0.27} \times 10^{48}$ & $-10.12$ \\
           & GC    & $4.45^{+0.58}_{-0.34}$ & $(3.93^{+0.38}_{-0.39}) \times 10^{2}$ & $2.30^{+0.80}_{-0.15} \times 10^{52}$ & $-4.84$ \\
    \tableline
    \multicolumn{6}{c}{Hadronic Origin (PL/ECPL)} \\
    \tableline
    Loop I & Local & $2.70^{+0.07}_{-0.06}$ & -- & $6.49^{+0.34}_{-0.30} \times 10^{49}$ & $-13.14$ \\
           & GC    &  &  & $2.58^{+0.14}_{-0.12} \times 10^{55}$ &  \\
    eB     & Local & $2.19^{+0.15}_{-0.17}$ & $0.10^{+0.15}_{-0.05}$ & $2.54^{+0.19}_{-0.22} \times 10^{49}$ & $-4.31$ \\
           & GC    &  &  & $1.02^{+0.09}_{-0.08} \times 10^{55}$ &  \\
    \tableline
  \end{tabular}
\end{table}

The primary systematic uncertainty in characterizing large-scale diffuse structures arises from the modeling of the DGE. Given that the faint southern residuals constitute only a small fraction of the total high-latitude flux, potential mis-modeling of interstellar gas or cosmic-ray distributions could, in principle, bias the derived properties. To address this, we performed a rigorous systematic test by employing an ensemble of 64 distinct DGE templates, covering a wide range of physical assumptions regarding the Galaxy's properties. Crucially, the southern structure is detected with high statistical significance across all 64 models, demonstrating that the identified excess is not an artifact of a specific background choice. While further refinements in DGE modeling will continue to improve spectral precision and may help better quantify the contribution of the large-scale excess relative to the diffuse background, the consistency across the tested DGE models suggests that the excess itself is unlikely to disappear.

\section{Discussion and Conclusion}\label{sec4}

{The Fermi bubbles provide a well-established example of large-scale gamma-ray emission associated with past activity of the Galactic Center. A less settled question concerns the broader high-latitude structures surrounding them, especially the NPS/Loop I system, whose distance and physical origin have long been debated between local superbubble and Galactic-scale outflow interpretations \citep{lallement2022northpolarspurloopi,yang2022fermi,sarkar2023misaligned,zhang2024magnetized}. The eBs provide a morphological bridge between the FBs and the debated NPS/Loop I system, with their large-scale bipolar X-ray structure encompassing the FBs and extending toward the NPS/Loop I region \citep{predehl2020detection,Liu2024}. This morphology strengthens the interpretation that these high-latitude structures are connected to a Galactic-center outflow rather than to isolated local foreground shells. On the gamma-ray side, a northern counterpart associated with the NPS/Loop I region has already been identified \citep{su2010giant}, whereas a comparable southern large-scale component has remained unclear. Within this framework, our detection of a large-scale southern gamma-ray structure provides a candidate high-energy counterpart to the southern eB region. Its morphology is better described by a filled eB-like template than by the adopted Wolleben Loop I shell geometry, and the resulting northern--southern configuration is more naturally described as a gamma-ray counterpart to the larger eB system surrounding the FBs. This morphology may motivate a Galactic-scale outflow interpretation, in which the FBs trace the bright inner gamma-ray lobes and the eBs delineate a larger, more extended structure. Under the eB-template interpretation, the substantially fainter and softer southern component further indicates a pronounced north--south asymmetry in the large-scale non-thermal emission. However, because the southern residual lies in a region affected by large-scale foreground structures, a localized foreground contribution cannot be decisively ruled out with the present data.}

Physically, this large-scale picture generally supports a Galactic-center feedback/outflow framework. For instance, \cite{yang2022fermi} proposed a unified leptonic-jet model where a single episode of jet activity from Sgr A* simultaneously inflated the FBs and the eBs. However, in their fiducial model, the gamma-ray emission is predominantly confined to the primary cosmic-ray electrons within the compact FBs, while the eBs merely trace the thermal bremsstrahlung from the outer forward shock. Consequently, their model lacks the capacity to predict a large-scale, diffuse gamma-ray structure extending to the eB boundaries. If associated with the eB system, this extended gamma-ray emission would provide a new constraint for theoretical models, implying the presence of a macroscopic non-thermal electron population at much larger radii than previously expected. The physical nature of these particles, however, is best decoded through their energy distribution. Recent simulations of double-episode jet activity further suggest that the eB boundaries themselves may be defined by a preceding forward shock that pre-accelerates electrons, providing a potential mechanism for such extended gamma-ray emission \citep{zhang2026doubleepisodejetgenesiserosita}.

{In a leptonic interpretation, the fact that the large-scale emission
is softer than the inner FBs naturally points to radiative aging and
transport effects of cosmic-ray electrons in the Galactic halo.
Adopting a representative field of \(B\sim5~\mu{\rm G}\),
motivated by magnetic-field estimates for the radio lobes
\citep{carretti2013giant}, the several-hundred-GeV to a few-TeV
electrons contributing to the LAT-band emission through inverse-Compton
scattering on the cosmic microwave background have synchrotron and
inverse-Compton cooling times of order \(0.1\)--\(1\)~Myr.
For conventional halo diffusion coefficients, a purely diffusive
supply from the pre-existing FB electron population is therefore
unlikely to fill the outer eB volume before substantial cooling occurs.
A Galactic-scale leptonic scenario instead requires rapid advective
transport in a large-scale outflow and/or distributed in situ
reacceleration in the outer bubble region.}

{The north--south spectral and brightness asymmetry may then reflect differences in transport time, radiative or adiabatic losses, and shock properties. Here, in situ acceleration or reacceleration should be understood as local particle energization within the outer bubble region, including the forward shock and the kpc-scale downstream shocked plasma, rather than as a purely diffusive supply from the pre-existing FB electron population. If in situ acceleration or reacceleration is important, the softer southern spectrum may indicate an older or weaker shock, or a shock propagating into a hotter halo with a higher sound speed, leading to a lower effective Mach number. The lower southern surface brightness may further reflect a smaller energy flux available for particle acceleration, a lower density of accelerated electrons, or a reduced acceleration efficiency. A north--south difference in the circumgalactic gas density
can itself produce substantial asymmetries in the X-ray and gamma-ray
surface brightness in star-formation-driven wind models
\citep{sarkar2019possible}. This interpretation is qualitatively consistent with hydrodynamic simulations in which a dynamic circumgalactic-medium wind disturbs the Galactic halo and redistributes its gas density and metallicity, thereby helping to explain the asymmetric morphology and surface brightness of the eROSITA bubbles \citep{mou2023asymmetric}. Such an asymmetric halo environment provides a natural context for north-south differences in shock evolution, particle acceleration, and radiative aging.
}

Beyond the leptonic interpretations, a hadronic origin for the observed gamma-ray emission warrants careful consideration. In this scenario, cosmic-ray protons interacting with the ambient thermal gas in the Galactic halo would produce gamma-rays via $\pi^0$ decay. Theoretically, such a hadronic component would be less sensitive to the cooling losses that affect electrons, potentially offering a more straightforward explanation for the survival of high-energy particles at such large radii. Furthermore, a hadronic framework naturally predicts a correlated flux of high-energy neutrinos originating from the decay of charged pions produced in the same $pp$ interactions, although the observable neutrino energy range depends directly on the parent proton spectrum \citep{PhysRevD.89.103003,ahlers2014pinpointing}.

{
For the Galactic-scale eB interpretation with a halo gas density of
\(n_{\rm H}\simeq10^{-3}~{\rm cm^{-3}}\), consistent with
observational and modeling estimates
\citep{miller2015constraining,faerman2022exploring}, the southern
eB-like component requires a cosmic-ray proton energy of
\(W_p \sim 10^{55}
(10^{-3}~{\rm cm^{-3}}/n_{\rm H})
(d/8~{\rm kpc})^2\)~erg.
Assuming the same effective target density in both hemispheres
and accounting for the approximately seven times more luminous
northern counterpart, the total proton energy for the full bipolar
system approaches
\(W_p\sim10^{56}
(10^{-3}~{\rm cm^{-3}}/n_{\rm H})
(d/8~{\rm kpc})^2\)~erg.
Assuming a characteristic cosmic-ray acceleration efficiency of
\(\eta_{\rm CR}\sim0.1\), the
required mechanical energy approaches
\(E_{\rm mech}\sim10^{57}
(10^{-3}~{\rm cm^{-3}}/n_{\rm H})
(0.1/\eta_{\rm CR})
(d/8~{\rm kpc})^2\)~erg.
This is substantially higher than the
\(\sim10^{56}\)~erg energy scale estimated for the eROSITA bubbles
\citep{predehl2020detection}, making a hadronic interpretation
energetically demanding.
In contrast, for the same Galactic-scale interpretation, an
inverse-Compton leptonic interpretation requires a substantially
smaller non-thermal particle energy, although the exact budget depends
on the adopted radiation field, magnetic field, geometry, and
low-energy cutoff. Because both emission mechanisms can trace the
large-scale outflow structure, the present morphology and energetics
alone cannot uniquely discriminate between them. We therefore consider
a hadronic contribution energetically demanding, while noting
that a leptonic-dominated interpretation remains more economical in
terms of particle energy. Because the eB hadronic ECPL fit prefers a
relatively low proton cutoff, \(E_{c,p}\sim0.1~{\rm TeV}\), the
associated TeV neutrino signal would be strongly suppressed. Future
neutrino observations could test a hadronic contribution only if the
proton spectrum extends to sufficiently high energies or if an
additional harder hadronic component is present.
}

{In conclusion, we report the identification and characterization of a large-scale southern gamma-ray structure that is spatially consistent with the southern eROSITA bubble. As a plausible gamma-ray counterpart of the southern eB, it helps address the long-standing north-south asymmetry in the high-latitude gamma-ray sky.} While the spatial alignment and large-scale morphology are naturally interpreted within a Galactic-scale outflow scenario, our current data do not yet allow for a decisive rejection of localized interpretations. In this context, we have estimated the energy required for the relativistic particle population. For completeness, Appendix~\ref{apsed} also presents particle-energy estimates for a local-distance scenario, while the main discussion focuses on the Galactic-scale interpretation motivated by the eB-like morphology. Under the Galactic-scale interpretation, the southern structure would represent a significant energetic component of the Galactic halo and would impose new constraints on the energetics and duty cycle of Galactic nuclear feedback.

Ultimately, resolving the true nature of this emission, whether as a macroscopic tracer of the Milky Way’s past activity or a nearby feature, will require more refined multi-messenger observations. Higher-sensitivity data across multiple wavebands will be essential to further constrain the particle transport and disentangle the underlying leptonic and hadronic processes, providing a definitive understanding of these expansive structures within the Galactic ecosystem.

\begin{acknowledgments}
Xiaoyuan Huang is supported by the National Key Research and Development Program of China under grant 2022YFF0503304, the National Natural Science Foundation of China under grant 12322302, the Project for Young Scientists in Basic Research of Chinese Academy of Sciences under grant YSBR-061, and the Chinese Academy of Sciences. Bing Liu is supported by the Natural Science Foundation for General Program of Jiangsu Province of China under grant NO. BK20252108. Ruizhi Yang is supported by the NSFC under grant 12393854, 12588101, and by the natural science funding of Sichuan Province under grant 2025ZNSFSC0065. Ruizhi Yang gratefully acknowledges the support of Cyrus Chun Ying Tang Foundations and of the studio of Academician Zhao Zhengguo, Deep Space Exploration Laboratory. 

\end{acknowledgments}

\begin{contribution}
%%This section gives authors the space to recognize author contributions. The text inside this environment is NOT counted towards the total word quanta. At a minimum, manuscripts are expected to include this text:
Z. Xie performed the data analysis and contributed to the discussion of the results. X.-Y. Huang (corresponding author) conceived the study and developed the data processing pipeline. B. Liu provided the details and technical expertise for the SED analysis. R.-Z. Yang (corresponding author) conceived the study, conducted the theoretical interpretation, and oversaw the entire project. All authors participated in the preparation and final review of the manuscript.

%% But authors are expected to provide more specific details, e.g. 
%%
%%SC was responsible for writing and submitting the manuscript.
%%WWM came up with the initial research concept and edited the manuscript.
%%OTS obtained the funding and edited the manuscript.
%%EBF provided the formal analysis and validation. He also edited the manuscript.
%%GEH Supervised the undergraduates, wrote the software and administers the project github and Zenodo repositories.
%%
%% Authors can use the Contributor Role Taxonomy (CRediT) at
%% for ideas on how wroite a good statement tailored to their needs.

\end{contribution}

\facilities{Fermi-LAT}

\software{astropy \citep{astropy:2013, astropy:2018, astropy:2022}, HEALPix \citep{2005ApJ...622..759G}, GALPROP \citep{Porter_2022}, naima \citep{naima}.}


\appendix

\section{Background Model}\label{ap:bg}
To evaluate the systematic uncertainties associated with the DGE
background, we perform an extensive comparison across a library of
emission templates. The template set spans variations in the CR
source distribution, Galactic halo height, and interstellar gas
distribution. The resulting ensemble samples the associated
uncertainties in the predicted inverse-Compton, bremsstrahlung, and
\(\pi^0\)-decay emission, allowing us to quantify the systematic
uncertainty arising from diffuse foreground modeling.

Accurate characterization of the morphology and spectral features of the observed diffuse structures hinges critically on the reliable modeling of the Galactic diffuse background. However, inherent uncertainties in CR source distributions and the interstellar medium properties introduce potential biases. To systematically evaluate these uncertainties, we use the GALPROP diffuse-emission template set developed in previous Galactic diffuse gamma-ray studies~\citep{ackermann2012fermi}. In particular, the template set spans variations in the cosmic-ray halo height, source distribution, and gas density profile. Following the selection procedure adopted in Ref.~\citep{yang2014fermi}, models with halo heights $z=8~{\rm kpc}$ and $z=10~{\rm kpc}$ that fail to reproduce the $^{9}{\rm Be}/^{10}{\rm Be}$ data are discarded, and the remaining 64 GALPROP models are retained here as diffuse-emission templates.

\begin{figure}[htbp]
    \centering
    \includegraphics[width=0.5\linewidth]{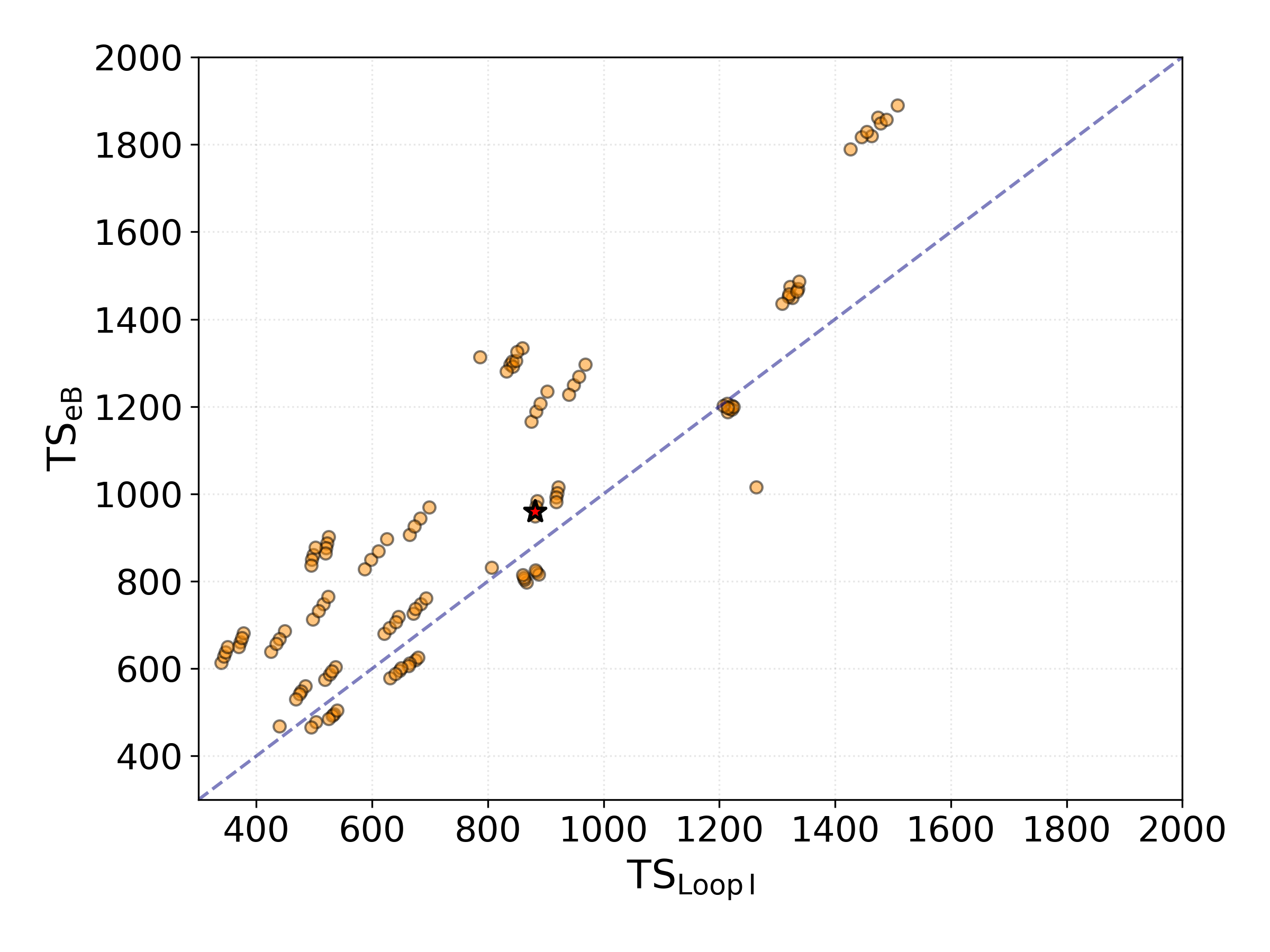}
    \includegraphics[width=1.0\linewidth]{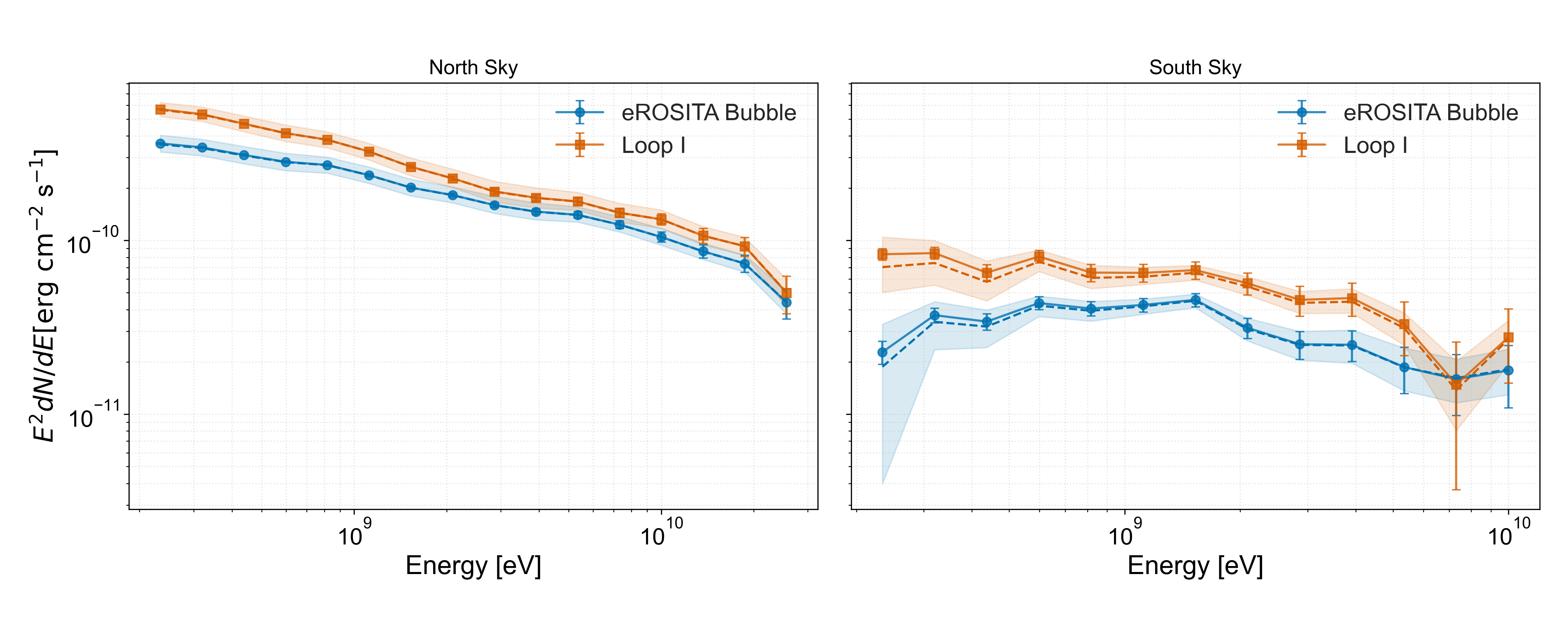}
    \caption{
    Top: Comparison of cumulative template TS values for the eB and Loop I templates across the 64 GALPROP diffuse-emission templates. Each value is computed relative to the corresponding baseline model without the additional large-scale template. The benchmark model is highlighted by the red star, and the navy dashed line indicates equal likelihood improvement for the two non-nested spatial hypotheses.\\
    Bottom: SEDs obtained using the 64 GALPROP diffuse-emission templates for the northern (left panel) and southern (right panel) large-scale structures. Dashed lines indicate the ensemble mean, and shaded regions denote the $1\sigma$ spread across the diffuse-emission templates. The data points show the SEDs obtained with the benchmark background model, with circles and squares denoting the eB and Loop I templates, respectively.}
    \label{fig:nscp}
\end{figure}

We investigate an ensemble of 64 GALPROP models to account for the systematic errors arising from the diffuse gamma-ray background. These models explore a diverse parameter space, including variations in CR halo height, source distributions, and gas density profiles. From this ensemble, we designate the template set that most closely aligns with the ensemble mean as our benchmark model, used for the benchmark SED points in Figs.~\ref{fig:all-spec} and \ref{fig:nscp}. This model serves as the representative baseline for our primary analysis, reflecting the central trend of the explored parameter space.

The resulting statistical and spectral variations are illustrated in Fig. \ref{fig:nscp}. By comparing the TS scores across the 64-model ensemble, we evaluate the fitting performance of different propagation scenarios. The corresponding SED comparisons reveal the impact of these parameter variations on the flux levels in both hemispheres, providing a robust quantification of the systematic uncertainties inherent in our SED characterization.

The benchmark background model is constructed by incorporating several physical and observational components. The DGE, comprising neutral pion decay, bremsstrahlung, and inverse-Compton scattering, is modeled using templates generated by the GALPROP code \citep{Porter_2022}. In this framework, the spatial distribution of CR sources is parameterized based on the pulsar distribution in the Galaxy \citep{lorimer2006parkes}. The diffusion coefficient for CR propagation in the interstellar medium is set to $D_{xx} = 8.1 \times 10^{28} \, \mathrm{cm^2 \, s^{-1}}$ at a reference rigidity of $4\,\mathrm{GV}$ with a scaling index of $0.33$. The model incorporates CR reacceleration with a constant Alfvén velocity of $v_A = 30 \, \mathrm{km \, s^{-1}}$ and a Galactic magnetic field with a local random field strength of $7.5 \, \mu\mathrm{G}$ within a CR halo of $z_h = 6\,\mathrm{kpc}$. The injection spectra for both nucleons and electrons are modeled as doubly broken power laws in rigidity; specifically, the proton injection index is $1.9$ below $11\,\mathrm{GV}$, $2.45$ above $11\,\mathrm{GV}$. 

\section{SED analysis}\label{apsed}
The derived energy spectrum of the detected large-scale structure is presented in Figure~\ref{fig:all-spec}. We perform a spectral fit using a log-parabola model, which reveals a softer spectrum compared to the canonical \textit{Fermi} bubbles. Specifically, the spectral index is notably larger (softer) across the observed energy range, indicating a depletion of high-energy CRs at larger Galactic latitudes. 

This spectral softening provides useful insight into the particle population responsible for the emission. Under a Galactic-scale leptonic interpretation, the observed spectral softening may be explained by outward transport of particles associated with the larger FB/eB system, together with radiative and adiabatic losses during expansion into the Galactic halo. Energy-dependent transport and less efficient reacceleration in the outer bubble region may further suppress the highest-energy electron population, naturally leading to a softer gamma-ray spectrum than that of the compact inner FBs.

To investigate the radiation mechanisms of the detected southern structure, we extract its SED and perform numerical modeling using the naima package \citep{naima}. We consider two primary distance scenarios that reflect the ongoing debate regarding the nature of the Loop I complex and its southern extension: a Local scenario ($d \approx 400$~pc) and a GC scenario ($d \approx 8$~kpc).

The physical environment, particularly the Interstellar Radiation Field (ISRF) that serves as the seed for inverse-Compton scattering, differs significantly between these two frameworks. In the GC scenario, where the structure is assumed to be associated with past Galactic Center activity and the FBs, its average altitude above the Galactic plane exceeds 4~kpc. At such heights, the stellar and infrared components of the ISRF are subdominant; consequently, our leptonic modeling for this scenario only accounts for the CMB as the target photon field. Conversely, in the Local scenario, the structure resides within the solar neighborhood ($r \approx 8.0$~kpc, $z \approx 0.2$~kpc). For this case, we adopt the detailed ISRF model from \citet{popescu2017radiation} to account for the significantly higher local photon density from starlight and dust re-emission.
\begin{figure}[htbp]
    \centering
     \includegraphics[width=0.48\linewidth]{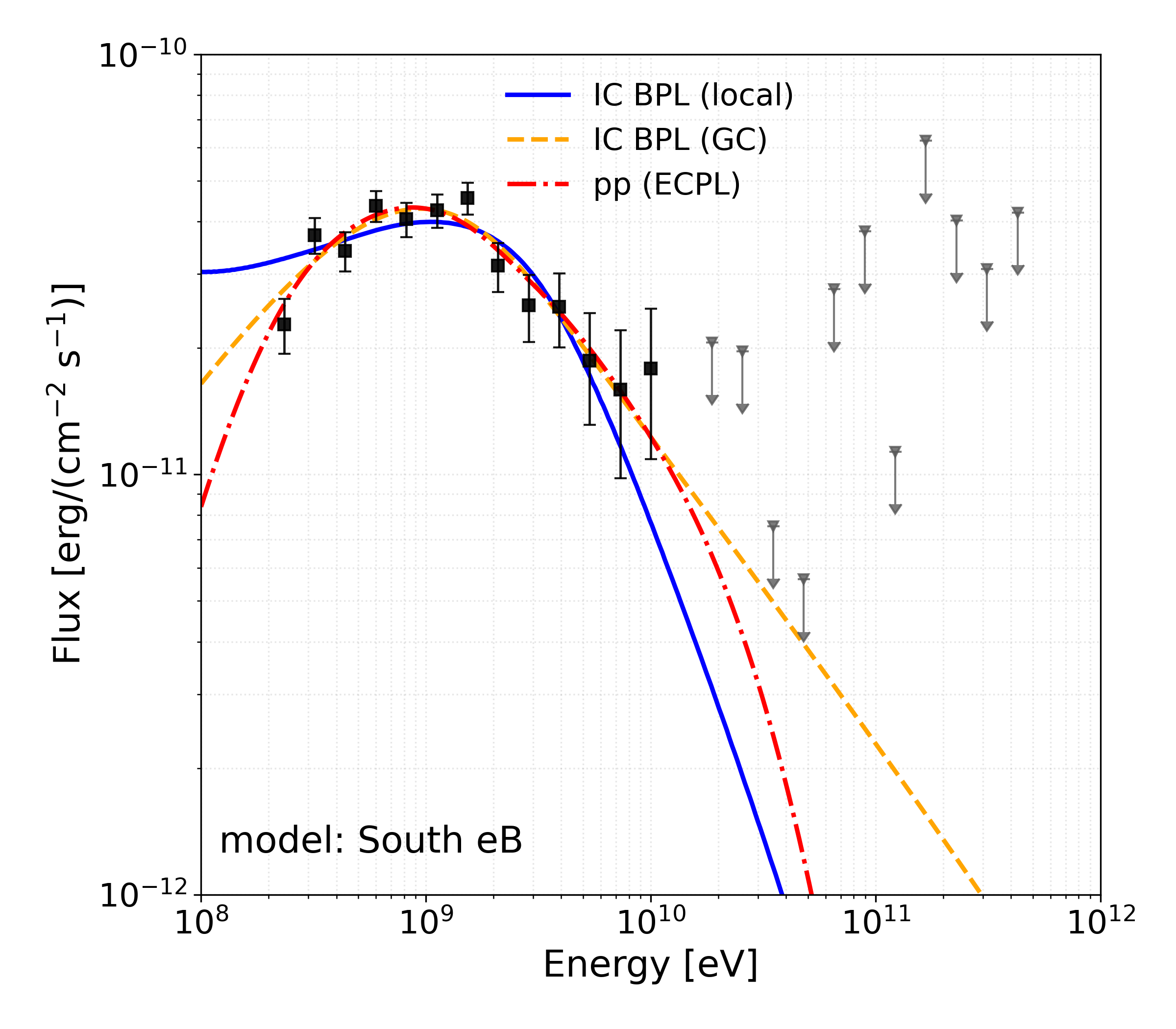}
     \includegraphics[width=0.48\linewidth]{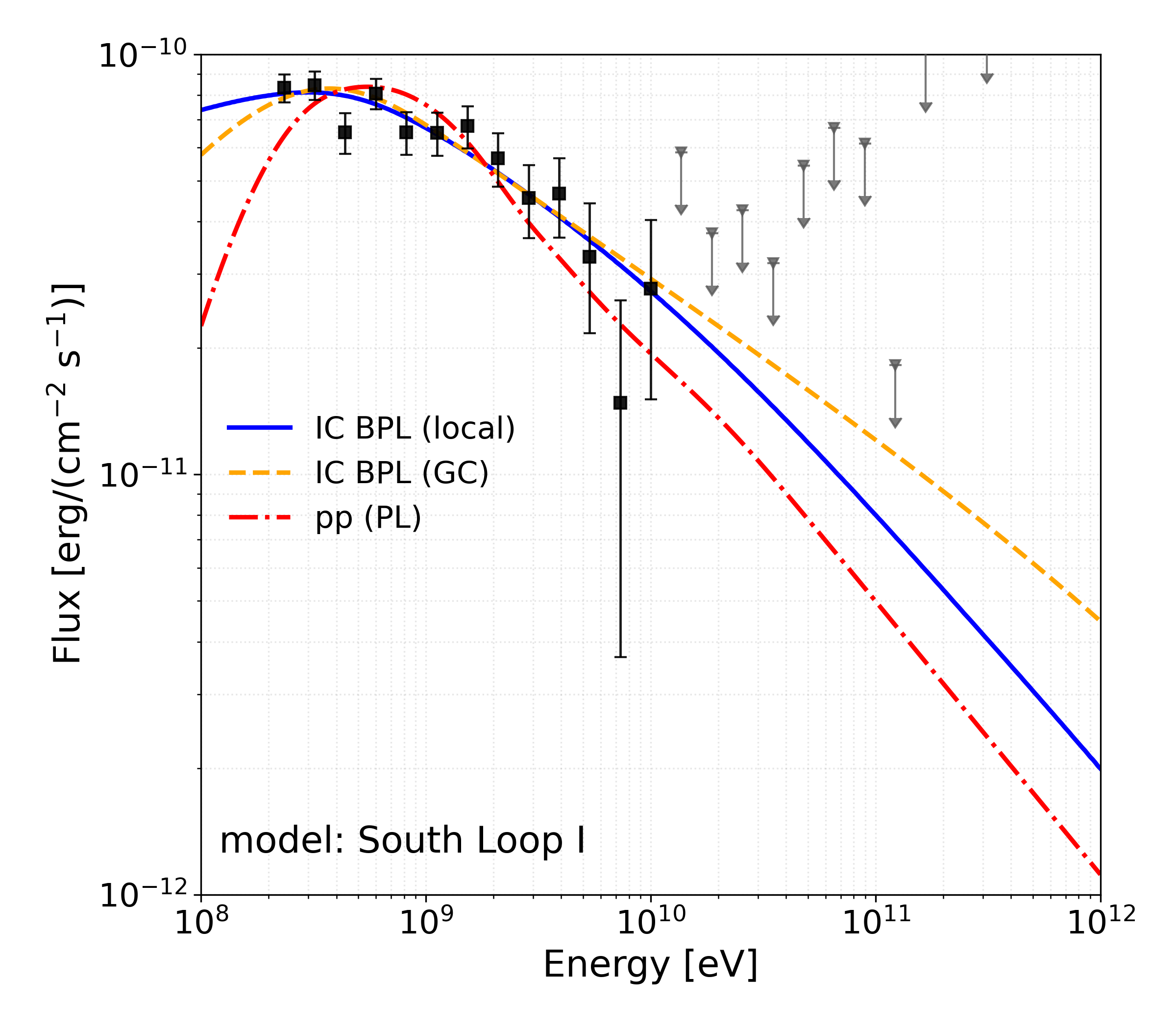}
    \caption{Spectral energy distribution of the southern structure fitted with leptonic and hadronic models. The left and right panels show the SEDs derived with the
southern eB and Loop I spatial templates, respectively. The leptonic scenario (inverse-Compton) assumes a BPL ($\gamma_1=1.5$) electron distribution. For the hadronic (pp) fits, a PL or ECPL proton distribution is adopted, with target gas densities assumed to be $n_{\text{H}} = 1.0 \text{ cm}^{-3}$ for the local components and $n_{\text{H}} = 10^{-3} \text{ cm}^{-3}$ for the Galactic-scale structure toward the GC. Upper limits are shown for reference and are not used in computing the model likelihoods reported in Table~\ref{tab:combined_fit}.}
    \label{pp}
\end{figure}

As shown in Fig. \ref{pp}, we evaluate both leptonic and hadronic populations within these environmental constraints. In the leptonic model, we adopt a broken power-law (BPL) electron distribution. {Because the present LAT SED alone does not tightly constrain the low-energy electron slope, we fix the low-energy index to $\gamma_1=1.5$ as a hard-spectrum phenomenological choice and fit the electron normalization, high-energy slope, and break energy.} For a given present-day electron distribution and target photon
field, the inverse-Compton spectrum does not depend on the magnetic
field. Since neither synchrotron data nor a time-dependent cooling model
is included, the magnetic field does not enter the present
\texttt{naima} likelihood fit. A higher field shortens the electron
cooling time but does not change the fitted inverse-Compton spectrum,
electron parameters, or likelihood. {For the hadronic ($pp$) scenario, we test both a power-law (PL) and an exponentially cutoff power-law (ECPL) proton distribution using \texttt{naima}. For the Loop I model, the detected SED points do not require the additional cutoff parameter, and we therefore adopt the simpler PL description. For the eB model, the ECPL form provides a useful phenomenological description of the detected SED while avoiding an excessive high-energy extrapolation when the upper limits are considered qualitatively. The high-energy upper limits are used only as a qualitative guide, and the likelihood values reported in Table~\ref{tab:combined_fit} are computed using only the detected data points.}

The inverse-Compton models show some deviations from individual low- and high-energy bins for both the eB and Loop I template SEDs. For the Loop I template, the local and GC models yield $\ln\mathcal{L}=-4.89$ and $-5.72$, respectively, and the small difference of $\Delta\ln\mathcal{L}=0.83$ does not indicate a clear preference. For the eB template, the GC model provides a better fit, with $\ln\mathcal{L}=-4.84$ compared with $-10.12$ for the local model. However, the local eB fit is based on a restricted BPL electron distribution with the low-energy index fixed at $\gamma_1=1.5$. We therefore cannot rule out a local inverse-Compton interpretation based on these fits alone.

Considering the spatial variation of the target gas, we adopt a target density of $n_{\mathrm{H}} = 1.0\,\mathrm{cm}^{-3}$ for the local interstellar medium, while $n_{\mathrm{H}} = 10^{-3}\,\mathrm{cm}^{-3}$ is adopted for the hot high-latitude Galactic halo in the GC-distance interpretation, consistent with current halo-gas estimates \citep{miller2015constraining,faerman2022exploring}.

For the southern component in the eB-template situation, the integrated photon flux over the unmasked portion of the spatial template used in the likelihood analysis is $1.0 \times 10^{-7}\ \mathrm{ph\ cm^{-2}\ s^{-1}}$. The inferred gamma-ray luminosity is $L_{\gamma} \approx 2.4 \times 10^{33}\ \mathrm{erg\ s^{-1}}$ assuming a local origin at $d = 400\ \mathrm{pc}$, and $L_{\gamma} \approx 9.7 \times 10^{35}\ \mathrm{erg\ s^{-1}}$ for a Galactic-scale distance of $d = 8\ \mathrm{kpc}$.
Under the same eB model framework, the northern component exhibits a significantly higher luminosity, with $L_{\gamma} \approx 1.8 \times 10^{34}\ \mathrm{erg\ s^{-1}}$ at $0.4\ \mathrm{kpc}$ and $L_{\gamma} \approx 7.1 \times 10^{36}\ \mathrm{erg\ s^{-1}}$ at $8\ \mathrm{kpc}$. This suggests that the northern structure is intrinsically brighter than the southern one, showing a clear north–south asymmetry, consistent with that seen in the eROSITA X-ray morphology.
For comparison, adopting the Loop I model leads to systematically higher gamma-ray luminosities. The northern component yields $L_{\gamma} \approx 2.3 \times 10^{34}\ \mathrm{erg\ s^{-1}}$ (0.4 kpc) and $9.2 \times 10^{36}\ \mathrm{erg\ s^{-1}}$ (8 kpc), while the southern component gives $4.5 \times 10^{33}\ \mathrm{erg\ s^{-1}}$ and $1.8 \times 10^{36}\ \mathrm{erg\ s^{-1}}$, respectively.

%% For this sample we use BibTeX plus aasjournalv7.bst to generate the

\bibliography{main}{}

@article{lorimer2006parkes,
  title={The Parkes Multibeam Pulsar Survey--VI. Discovery and timing of 142 pulsars and a Galactic population analysis},
  author={Lorimer, Duncan R and Faulkner, AJ and Lyne, AG and Manchester, Richard N and Kramer, M and McLaughlin, MA and Hobbs, G and Possenti, A and Stairs, IH and Camilo, F and others},
  journal={Monthly Notices of the Royal Astronomical Society},
  volume={372},
  number={2},
  pages={777--800},
  year={2006},
  publisher={Blackwell Publishing Ltd Oxford, UK}
}

@article{sarkar2024fermi,
  title={The Fermi/eROSITA bubbles: a look into the nuclear outflow from the Milky Way},
  author={Sarkar, Kartick C},
  journal={arXiv preprint arXiv:2403.09824},
  year={2024}
}

@article{yang2022fermi,
  title={Fermi and eROSITA bubbles as relics of the past activity of the Galaxy’s central black hole},
  author={Yang, H-Y Karen and Ruszkowski, Mateusz and Zweibel, Ellen G},
  journal={Nature Astronomy},
  volume={6},
  number={5},
  pages={584--591},
  year={2022},
  publisher={Nature Publishing Group UK London}
}

@article{yang2014fermi,
  title={The Fermi bubbles revisited},
  author={Yang, Rui-zhi and Aharonian, Felix and Crocker, Roland},
  journal={Astronomy \& Astrophysics},
  volume={567},
  pages={A19},
  year={2014},
  publisher={EDP Sciences}
}

@article{crocker2011fermi,
  title={Fermi bubbles: giant, multibillion-year-old reservoirs of galactic center cosmic rays},
  author={Crocker, Roland M and Aharonian, Felix},
  journal={Physical Review Letters},
  volume={106},
  number={10},
  pages={101102},
  year={2011},
  publisher={APS}
}

@article{guo2012fermi,
  title={The Fermi bubbles. I. Possible evidence for recent AGN jet activity in the galaxy},
  author={Guo, Fulai and Mathews, William G},
  journal={The Astrophysical Journal},
  volume={756},
  number={2},
  pages={181},
  year={2012},
  publisher={IOP Publishing}
}

@article{kataoka2018x,
  title={X-ray and gamma-ray observations of the Fermi bubbles and NPS/Loop I structures},
  author={Kataoka, Jun and Sofue, Yoshiaki and Inoue, Yoshiyuki and Akita, Masahiro and Nakashima, Shinya and Totani, Tomonori},
  journal={Galaxies},
  volume={6},
  number={1},
  pages={27},
  year={2018},
  publisher={MDPI}
}

@article{su2010giant,
  title={Giant gamma-ray bubbles from Fermi-LAT: active galactic nucleus activity or bipolar galactic wind?},
  author={Su, Meng and Slatyer, Tracy R and Finkbeiner, Douglas P},
  journal={The Astrophysical Journal},
  volume={724},
  number={2},
  pages={1044},
  year={2010},
  publisher={IOP Publishing}
}

@article{haslam1982408,
  title={A 408 MHz all-sky continuum survey. II-The atlas of contour maps},
  author={Haslam, CGT and Salter, CJ and Stoffel, H and Wilson, WEz},
  journal={Astronomy and Astrophysics Supplement Series, vol. 47, Jan. 1982, p. 1, 2, 4-51, 53-142.},
  volume={47},
  pages={1},
  year={1982}
}

@article{predehl2020detection,
  title={Detection of large-scale X-ray bubbles in the Milky Way halo},
  author={Predehl, P and Sunyaev, RA and Becker, W and Brunner, H and Burenin, R and Bykov, A and Cherepashchuk, A and Chugai, N and Churazov, E and Doroshenko, V and others},
  journal={Nature},
  volume={588},
  number={7837},
  pages={227--231},
  year={2020},
  publisher={Nature Publishing Group UK London}
}

@article{wolleben2007new,
  title={A new model for the Loop I (North Polar Spur) region},
  author={Wolleben, M},
  journal={The Astrophysical Journal},
  volume={664},
  number={1},
  pages={349},
  year={2007},
  publisher={IOP Publishing}
}

@ARTICLE{Liu2024,
       author = {{Liu}, Teng and {Merloni}, Andrea and {Sanders}, Jeremy and others},
        title = "{Morphological Evidence for the eROSITA Bubbles Being Giant and Distant Structures}",
      journal = {The Astrophysical Journal Letters},
         year = 2024,
       volume = {967},
       number = {2},
        pages = {L27},
          doi = {10.3847/2041-8213/ad47e0},
}

@article{popescu2017radiation,
  title={A radiation transfer model for the Milky Way: I. Radiation fields and application to high-energy astrophysics},
  author={Popescu, CC and Yang, R and Tuffs, RJ and Natale, Giovanni and Rushton, M and Aharonian, F},
  journal={Monthly Notices of the Royal Astronomical Society},
  volume={470},
  number={3},
  pages={2539--2558},
  year={2017},
  publisher={Oxford University Press}
}

@article{Porter_2022,
   title={The GALPROP Cosmic-ray Propagation and Nonthermal Emissions Framework: Release v57},
   volume={262},
   ISSN={1538-4365},
   url={http://dx.doi.org/10.3847/1538-4365/ac80f6},
   DOI={10.3847/1538-4365/ac80f6},
   number={1},
   journal={The Astrophysical Journal Supplement Series},
   publisher={American Astronomical Society},
   author={Porter, T. A. and Jóhannesson, G. and Moskalenko, I. V.},
   year={2022},
   month=sep, pages={30} }

@ARTICLE{naima,
   author = {{Zabalza}, V.},
    title = {naima: a Python package for inference of relativistic particle
             energy distributions from observed nonthermal spectra},
     year = 2015,
  journal = {Proc.~of International Cosmic Ray Conference 2015},
    pages = "922",
   eprint = {1509.03319},
   adsurl = {http://adsabs.harvard.edu/abs/2015arXiv150903319Z},
}

@misc{ballet2024fermilargeareatelescope,
      title={Fermi Large Area Telescope Fourth Source Catalog Data Release 4 (4FGL-DR4)}, 
      author={J. Ballet and P. Bruel and T. H. Burnett and B. Lott and The Fermi-LAT collaboration},
      year={2024},
      eprint={2307.12546},
      archivePrefix={arXiv},
      primaryClass={astro-ph.HE},
      url={https://arxiv.org/abs/2307.12546}, 
}

@article{Abdollahi_2022,
   title={Incremental Fermi Large Area Telescope Fourth Source Catalog},
   volume={260},
   ISSN={1538-4365},
   url={http://dx.doi.org/10.3847/1538-4365/ac6751},
   DOI={10.3847/1538-4365/ac6751},
   number={2},
   journal={The Astrophysical Journal Supplement Series},
   publisher={American Astronomical Society},
   author={Abdollahi, S. and Acero, F. and Baldini, L. and Ballet, J. and Bastieri, D. and Bellazzini, R. and Berenji, B. and Berretta, A. and Bissaldi, E. and Blandford, R. D. and Bloom, E. and Bonino, R. and Brill, A. and Britto, R. J. and Bruel, P. and Burnett, T. H. and Buson, S. and Cameron, R. A. and Caputo, R. and Caraveo, P. A. and Castro, D. and Chaty, S. and Cheung, C. C. and Chiaro, G. and Cibrario, N. and Ciprini, S. and Coronado-Blázquez, J. and Crnogorcevic, M. and Cutini, S. and D’Ammando, F. and De Gaetano, S. and Digel, S. W. and Di Lalla, N. and Dirirsa, F. and Di Venere, L. and Domínguez, A. and Fallah Ramazani, V. and Fegan, S. J. and Ferrara, E. C. and Fiori, A. and Fleischhack, H. and Franckowiak, A. and Fukazawa, Y. and Funk, S. and Fusco, P. and Galanti, G. and Gammaldi, V. and Gargano, F. and Garrappa, S. and Gasparrini, D. and Giacchino, F. and Giglietto, N. and Giordano, F. and Giroletti, M. and Glanzman, T. and Green, D. and Grenier, I. A. and Grondin, M.-H. and Guillemot, L. and Guiriec, S. and Gustafsson, M. and Harding, A. K. and Hays, E. and Hewitt, J. W. and Horan, D. and Hou, X. and Jóhannesson, G. and Karwin, C. and Kayanoki, T. and Kerr, M. and Kuss, M. and Landriu, D. and Larsson, S. and Latronico, L. and Lemoine-Goumard, M. and Li, J. and Liodakis, I. and Longo, F. and Loparco, F. and Lott, B. and Lubrano, P. and Maldera, S. and Malyshev, D. and Manfreda, A. and Martí-Devesa, G. and Mazziotta, M. N. and Mereu, I. and Meyer, M. and Michelson, P. F. and Mirabal, N. and Mitthumsiri, W. and Mizuno, T. and Moiseev, A. A. and Monzani, M. E. and Morselli, A. and Moskalenko, I. V. and Negro, M. and Nuss, E. and Omodei, N. and Orienti, M. and Orlando, E. and Paneque, D. and Pei, Z. and Perkins, J. S. and Persic, M. and Pesce-Rollins, M. and Petrosian, V. and Pillera, R. and Poon, H. and Porter, T. A. and Principe, G. and Rainò, S. and Rando, R. and Rani, B. and Razzano, M. and Razzaque, S. and Reimer, A. and Reimer, O. and Reposeur, T. and Sánchez-Conde, M. and Saz Parkinson, P. M. and Scotton, L. and Serini, D. and Sgrò, C. and Siskind, E. J. and Smith, D. A. and Spandre, G. and Spinelli, P. and Sueoka, K. and Suson, D. J. and Tajima, H. and Tak, D. and Thayer, J. B. and Thompson, D. J. and Torres, D. F. and Troja, E. and Valverde, J. and Wood, K. and Zaharijas, G.},
   year={2022},
   month=jun, pages={53} }

@ARTICLE{2005ApJ...622..759G,
       author = {{G{\'o}rski}, K.~M. and {Hivon}, E. and {Banday}, A.~J. and {Wandelt}, B.~D. and {Hansen}, F.~K. and {Reinecke}, M. and {Bartelmann}, M.},
        title = "{HEALPix: A Framework for High-Resolution Discretization and Fast Analysis of Data Distributed on the Sphere}",
      journal = {\apj},
         year = 2005,
        month = apr,
       volume = {622},
       number = {2},
        pages = {759-771},
          doi = {10.1086/427976},
archivePrefix = {arXiv},
       eprint = {astro-ph/0409513},
 primaryClass = {astro-ph},
       adsurl = {https://ui.adsabs.harvard.edu/abs/2005ApJ...622..759G}
}

@article{mou2023asymmetric,
  title={Asymmetric eROSITA bubbles as the evidence of a circumgalactic medium wind},
  author={Mou, Guobin and Sun, Dongze and Fang, Taotao and Wang, Wei and Zhang, Ruiyu and Yuan, Feng and Sofue, Yoshiaki and Wang, Tinggui and He, Zhicheng},
  journal={Nature Communications},
  volume={14},
  number={1},
  pages={781},
  year={2023},
  publisher={Nature Publishing Group UK London}
}

@misc{lallement2022northpolarspurloopi,
      title={North Polar Spur/Loop I: gigantic outskirt of the Northern Fermi bubble or nearby hot gas cavity blown by supernovae?}, 
      author={Rosine Lallement},
      year={2022},
      eprint={2203.01312},
      archivePrefix={arXiv},
      primaryClass={astro-ph.GA},
      url={https://arxiv.org/abs/2203.01312}, 
}

@article{PhysRevD.89.103003,
  title = {Galactic halo origin of the neutrinos detected by IceCube},
  author = {Taylor, Andrew M. and Gabici, Stefano and Aharonian, Felix},
  journal = {Phys. Rev. D},
  volume = {89},
  issue = {10},
  pages = {103003},
  numpages = {7},
  year = {2014},
  month = {May},
  publisher = {American Physical Society},
  doi = {10.1103/PhysRevD.89.103003},
  url = {https://link.aps.org/doi/10.1103/PhysRevD.89.103003}
}

@article{ahlers2014pinpointing,
  title={Pinpointing extragalactic neutrino sources in light of recent IceCube observations},
  author={Ahlers, Markus and Halzen, Francis},
  journal={arXiv preprint arXiv:1406.2160},
  year={2014}
}

@article{ackermann2014spectrum,
  title={The spectrum and morphology of the Fermi bubbles},
  author={Ackermann, Markus and Albert, A and Atwood, WB and Baldini, Luca and Ballet, J and Barbiellini, G and Bastieri, Denis and Bellazzini, R and Bissaldi, Elisabetta and Blandford, RD and others},
  journal={The Astrophysical Journal},
  volume={793},
  number={1},
  pages={64},
  year={2014},
  publisher={The American Astronomical Society}
}

@article{bland2003large,
  title={The large-scale bipolar wind in the galactic center},
  author={Bland-Hawthorn, Joss and Cohen, Martin},
  journal={The Astrophysical Journal},
  volume={582},
  number={1},
  pages={246--256},
  year={2003}
}

@article{akita2018diffuse,
  title={Diffuse X-ray emission from the northern arc of Loop I observed with Suzaku},
  author={Akita, Masahiro and Kataoka, Jun and Arimoto, Makoto and Sofue, Yoshiaki and Totani, Tomonori and Inoue, Yoshiyuki and Nakashima, Shinya},
  journal={The Astrophysical Journal},
  volume={862},
  number={1},
  pages={88},
  year={2018},
  publisher={The American Astronomical Society}
}

@article{finkbeiner2004microwave,
  title={Microwave interstellar medium emission observed by the Wilkinson Microwave Anisotropy Probe},
  author={Finkbeiner, Douglas P},
  journal={The Astrophysical Journal},
  volume={614},
  number={1},
  pages={186--193},
  year={2004}
}

@article{bartlett2013planck,
  title={Planck intermediate results: IX. Detection of the galactic haze with planck},
  author={Bartlett, JG and Cardoso, JF and Delabrouille, J and Ganga, K and Piat, M and Rosset, C and Smoot, GF and L{\"a}hteenm{\"a}ki, A and Poutanen, T and Kunz, M and others},
  journal={Astronomy and Astrophysics},
  volume={554},
  pages={A139--A139},
  year={2013},
  publisher={Springer-Verlag GmbH}
}

@article{dobler2012last,
  title={A last look at the microwave haze/bubbles with WMAP},
  author={Dobler, Gregory},
  journal={The Astrophysical Journal},
  volume={750},
  number={1},
  pages={17},
  year={2012},
  publisher={The American Astronomical Society}
}

@article{sofue1979radio,
  title={Radio continuum observations of the North Polar Spur at 1420 MHz},
  author={Sofue, Yoshiaki and Reich, W},
  journal={Astronomy and Astrophysics Supplement Series, vol. 38, Nov. 1979, p. 251-263.},
  volume={38},
  pages={251--263},
  year={1979}
}

@article{snowden1997rosat,
  title={ROSAT survey diffuse X-ray background maps. II.},
  author={Snowden, SL and Egger, R and Freyberg, MJ and McCammon, D and Plucinsky, PP and Sanders, WT and Schmitt, JHMM and Tr{\"u}mper, J and Voges, W},
  journal={The Astrophysical Journal},
  volume={485},
  number={1},
  pages={125--135},
  year={1997}
}

@article{sarkar2023misaligned,
  title={Misaligned jets from Sgr A* and the origin of Fermi/eROSITA bubbles},
  author={Sarkar, Kartick C and Mondal, Santanu and Sharma, Prateek and Piran, Tsvi},
  journal={The Astrophysical Journal},
  volume={951},
  number={1},
  pages={36},
  year={2023},
  publisher={The American Astronomical Society}
}

@misc{zhang2026doubleepisodejetgenesiserosita,
      title={The Double-Episode Jet Genesis of the eROSITA and Fermi Bubbles}, 
      author={Ruiyu Zhang and Fulai Guo and Shaokun Xie and Ruofei Zhang and Shumin Wang and Guobin Mou and Xiaodong Duan},
      year={2026},
      eprint={2507.13665},
      archivePrefix={arXiv},
      primaryClass={astro-ph.HE},
      url={https://arxiv.org/abs/2507.13665}, 
}

@article{zhang2024magnetized,
  title={A magnetized Galactic halo from inner Galaxy outflows},
  author={Zhang, He-Shou and Ponti, Gabriele and Carretti, Ettore and Liu, Ruo-Yu and Morris, Mark R and Haverkorn, Marijke and Locatelli, Nicola and Zheng, Xueying and Aharonian, Felix and Zhang, Hai-Ming and others},
  journal={Nature Astronomy},
  volume={8},
  number={11},
  pages={1416--1428},
  year={2024},
  publisher={Nature Publishing Group UK London}
}

@misc{yeung2026srgerositadiffusesoftxray,
      title={The SRG/eROSITA diffuse soft X-ray background II. spectra and morphology of the eROSITA bubbles in the western Galactic hemisphere}, 
      author={Michael C. H. Yeung and Martin G. F. Mayer and Andy Strong and Michael J. Freyberg and Gabriele Ponti and Konrad Dennerl and Junjie Mao and Manami Sasaki and Xueying Zheng and Jeremy S. Sanders and Yi Zhang and Jiejia Liu and Liyi Gu and Werner Becker and Frank Haberl and Teng Liu and Andrea Merloni and Peter Predehl},
      year={2026},
      eprint={2605.02998},
      archivePrefix={arXiv},
      primaryClass={astro-ph.HE},
      url={https://arxiv.org/abs/2605.02998}, 
}

@article{Mertsch:2013pua,
    author = "Mertsch, Philipp and Sarkar, Subir",
    title = "{Loops and spurs: The angular power spectrum of the Galactic synchrotron background}",
    eprint = "1304.1078",
    archivePrefix = "arXiv",
    primaryClass = "astro-ph.GA",
    doi = "10.1088/1475-7516/2013/06/041",
    journal = "JCAP",
    volume = "06",
    pages = "041",
    year = "2013"
}

@article{Liu:2014mpa,
    author = "Liu, Hao and Mertsch, Philipp and Sarkar, Subir",
    title = "{Fingerprints of Galactic Loop I on the Cosmic Microwave Background}",
    eprint = "1404.1899",
    archivePrefix = "arXiv",
    primaryClass = "astro-ph.CO",
    doi = "10.1088/2041-8205/789/2/L29",
    journal = "Astrophys. J. Lett.",
    volume = "789",
    number = "2",
    pages = "L29",
    year = "2014"
}

@article{vonHausegger:2015flg,
    author = "von Hausegger, Sebastian and Liu, Hao and Mertsch, Philipp and Sarkar, Subir",
    title = "{Footprints of Loop I on Cosmic Microwave Background Maps}",
    eprint = "1511.08207",
    archivePrefix = "arXiv",
    primaryClass = "astro-ph.CO",
    doi = "10.1088/1475-7516/2016/03/023",
    journal = "JCAP",
    volume = "03",
    pages = "023",
    year = "2016"
}

@article{ackermann2012fermi,
  title={Fermi-LAT observations of the diffuse $\gamma$-ray emission: implications for cosmic rays and the interstellar medium},
  author={Ackermann, Markus and Ajello, Marco and Atwood, WB and Baldini, Luca and Ballet, Jean and Barbiellini, Guido and Bastieri, Denis and Bechtol, K and Bellazzini, R and Berenji, B and others},
  journal={The Astrophysical Journal},
  volume={750},
  number={1},
  pages={3},
  year={2012},
  publisher={The American Astronomical Society}
}

@article{miller2015constraining,
  title={Constraining the Milky Way's hot gas halo with O VII and O VIII emission lines},
  author={Miller, Matthew J and Bregman, Joel N},
  journal={The Astrophysical Journal},
  volume={800},
  number={1},
  pages={14},
  year={2015},
  publisher={The American Astronomical Society}
}

@article{faerman2022exploring,
  title={Exploring the Milky Way circumgalactic medium in a cosmological context with a semianalytic model},
  author={Faerman, Yakov and Pandya, Viraj and Somerville, Rachel S and Sternberg, Amiel},
  journal={The Astrophysical Journal},
  volume={928},
  number={1},
  pages={37},
  year={2022},
  publisher={The American Astronomical Society}
}

@article{carretti2013giant,
  title={Giant magnetized outflows from the centre of the Milky Way},
  author={Carretti, Ettore and Crocker, Roland M and Staveley-Smith, Lister and Haverkorn, Marijke and Purcell, Cormac and Gaensler, BM and Bernardi, Gianni and Kesteven, Michael J and Poppi, Sergio},
  journal={Nature},
  volume={493},
  number={7430},
  pages={66--69},
  year={2013},
  publisher={Nature Publishing Group UK London}
}

@article{astropy:2013,
        Adsurl = {https://adsabs.harvard.edu/abs/2013A%26A...558A..33A},
        Archiveprefix = {arXiv},
        Author = {{Astropy Collaboration} and {Robitaille}, T.~P. and {Tollerud}, E.~J. and {Greenfield}, P. and {Droettboom}, M. and {Bray}, E. and {Aldcroft}, T. and {Davis}, M. and {Ginsburg}, A. and {Price-Whelan}, A.~M. and {Kerzendorf}, W.~E. and {Conley}, A. and {Crighton}, N. and {Barbary}, K. and {Muna}, D. and {Ferguson}, H. and {Grollier}, F. and {Parikh}, M.~M. and {Nair}, P.~H. and {Unther}, H.~M. and {Deil}, C. and {Woillez}, J. and {Conseil}, S. and {Kramer}, R. and {Turner}, J.~E.~H. and {Singer}, L. and {Fox}, R. and {Weaver}, B.~A. and {Zabalza}, V. and {Edwards}, Z.~I. and {Azalee Bostroem}, K. and {Burke}, D.~J. and {Casey}, A.~R. and {Crawford}, S.~M. and {Dencheva}, N. and {Ely}, J. and {Jenness}, T. and {Labrie}, K. and {Lim}, P.~L. and {Pierfederici}, F. and {Pontzen}, A. and {Ptak}, A. and {Refsdal}, B. and {Servillat}, M. and {Streicher}, O.},
        Doi = {10.1051/0004-6361/201322068},
        Eid = {A33},
        Eprint = {1307.6212},
        Journal = {\aap},
        Month = oct,
        Pages = {A33},
        Primaryclass = {astro-ph.IM},
        Title = {{Astropy: A community Python package for astronomy}},
        Volume = 558,
        Year = 2013}

@ARTICLE{astropy:2018,
               author = {{Astropy Collaboration} and {Price-Whelan}, A.~M. and
                 {Sip{\H{o}}cz}, B.~M. and {G{\"u}nther}, H.~M. and {Lim}, P.~L. and
                 {Crawford}, S.~M. and {Conseil}, S. and {Shupe}, D.~L. and
                 {Craig}, M.~W. and {Dencheva}, N. and {Ginsburg}, A. and {Vand
                erPlas}, J.~T. and {Bradley}, L.~D. and {P{\'e}rez-Su{\'a}rez}, D. and
                 {de Val-Borro}, M. and {Aldcroft}, T.~L. and {Cruz}, K.~L. and
                 {Robitaille}, T.~P. and {Tollerud}, E.~J. and {Ardelean}, C. and
                 {Babej}, T. and {Bach}, Y.~P. and {Bachetti}, M. and {Bakanov}, A.~V. and
                 {Bamford}, S.~P. and {Barentsen}, G. and {Barmby}, P. and
                 {Baumbach}, A. and {Berry}, K.~L. and {Biscani}, F. and {Boquien}, M. and
                 {Bostroem}, K.~A. and {Bouma}, L.~G. and {Brammer}, G.~B. and
                 {Bray}, E.~M. and {Breytenbach}, H. and {Buddelmeijer}, H. and
                 {Burke}, D.~J. and {Calderone}, G. and {Cano Rodr{\'\i}guez}, J.~L. and
                 {Cara}, M. and {Cardoso}, J.~V.~M. and {Cheedella}, S. and {Copin}, Y. and
                 {Corrales}, L. and {Crichton}, D. and {D'Avella}, D. and {Deil}, C. and
                 {Depagne}, {\'E}. and {Dietrich}, J.~P. and {Donath}, A. and
                 {Droettboom}, M. and {Earl}, N. and {Erben}, T. and {Fabbro}, S. and
                 {Ferreira}, L.~A. and {Finethy}, T. and {Fox}, R.~T. and
                 {Garrison}, L.~H. and {Gibbons}, S.~L.~J. and {Goldstein}, D.~A. and
                 {Gommers}, R. and {Greco}, J.~P. and {Greenfield}, P. and
                 {Groener}, A.~M. and {Grollier}, F. and {Hagen}, A. and {Hirst}, P. and
                 {Homeier}, D. and {Horton}, A.~J. and {Hosseinzadeh}, G. and {Hu}, L. and
                 {Hunkeler}, J.~S. and {Ivezi{\'c}}, {\v{Z}}. and {Jain}, A. and
                 {Jenness}, T. and {Kanarek}, G. and {Kendrew}, S. and {Kern}, N.~S. and
                 {Kerzendorf}, W.~E. and {Khvalko}, A. and {King}, J. and {Kirkby}, D. and
                 {Kulkarni}, A.~M. and {Kumar}, A. and {Lee}, A. and {Lenz}, D. and
                 {Littlefair}, S.~P. and {Ma}, Z. and {Macleod}, D.~M. and
                 {Mastropietro}, M. and {McCully}, C. and {Montagnac}, S. and
                 {Morris}, B.~M. and {Mueller}, M. and {Mumford}, S.~J. and {Muna}, D. and
                 {Murphy}, N.~A. and {Nelson}, S. and {Nguyen}, G.~H. and
                 {Ninan}, J.~P. and {N{\"o}the}, M. and {Ogaz}, S. and {Oh}, S. and
                 {Parejko}, J.~K. and {Parley}, N. and {Pascual}, S. and {Patil}, R. and
                 {Patil}, A.~A. and {Plunkett}, A.~L. and {Prochaska}, J.~X. and
                 {Rastogi}, T. and {Reddy Janga}, V. and {Sabater}, J. and
                 {Sakurikar}, P. and {Seifert}, M. and {Sherbert}, L.~E. and
                 {Sherwood-Taylor}, H. and {Shih}, A.~Y. and {Sick}, J. and
                 {Silbiger}, M.~T. and {Singanamalla}, S. and {Singer}, L.~P. and
                 {Sladen}, P.~H. and {Sooley}, K.~A. and {Sornarajah}, S. and
                 {Streicher}, O. and {Teuben}, P. and {Thomas}, S.~W. and
                 {Tremblay}, G.~R. and {Turner}, J.~E.~H. and {Terr{\'o}n}, V. and
                 {van Kerkwijk}, M.~H. and {de la Vega}, A. and {Watkins}, L.~L. and
                 {Weaver}, B.~A. and {Whitmore}, J.~B. and {Woillez}, J. and
                 {Zabalza}, V. and {Astropy Contributors}},
                title = "{The Astropy Project: Building an Open-science Project and Status of the v2.0 Core Package}",
              journal = {\aj},
                 year = 2018,
                month = sep,
               volume = {156},
               number = {3},
                  eid = {123},
                pages = {123},
                  doi = {10.3847/1538-3881/aabc4f},
        archivePrefix = {arXiv},
               eprint = {1801.02634},
         primaryClass = {astro-ph.IM},
               adsurl = {https://ui.adsabs.harvard.edu/abs/2018AJ....156..123A}
        }

@ARTICLE{astropy:2022,
               author = {{Astropy Collaboration} and {Price-Whelan}, Adrian M. and {Lim}, Pey Lian and {Earl}, Nicholas and {Starkman}, Nathaniel and {Bradley}, Larry and {Shupe}, David L. and {Patil}, Aarya A. and {Corrales}, Lia and {Brasseur}, C.~E. and {N{"o}the}, Maximilian and {Donath}, Axel and {Tollerud}, Erik and {Morris}, Brett M. and {Ginsburg}, Adam and {Vaher}, Eero and {Weaver}, Benjamin A. and {Tocknell}, James and {Jamieson}, William and {van Kerkwijk}, Marten H. and {Robitaille}, Thomas P. and {Merry}, Bruce and {Bachetti}, Matteo and {G{"u}nther}, H. Moritz and {Aldcroft}, Thomas L. and {Alvarado-Montes}, Jaime A. and {Archibald}, Anne M. and {B{'o}di}, Attila and {Bapat}, Shreyas and {Barentsen}, Geert and {Baz{'a}n}, Juanjo and {Biswas}, Manish and {Boquien}, M{'e}d{'e}ric and {Burke}, D.~J. and {Cara}, Daria and {Cara}, Mihai and {Conroy}, Kyle E. and {Conseil}, Simon and {Craig}, Matthew W. and {Cross}, Robert M. and {Cruz}, Kelle L. and {D'Eugenio}, Francesco and {Dencheva}, Nadia and {Devillepoix}, Hadrien A.~R. and {Dietrich}, J{"o}rg P. and {Eigenbrot}, Arthur Davis and {Erben}, Thomas and {Ferreira}, Leonardo and {Foreman-Mackey}, Daniel and {Fox}, Ryan and {Freij}, Nabil and {Garg}, Suyog and {Geda}, Robel and {Glattly}, Lauren and {Gondhalekar}, Yash and {Gordon}, Karl D. and {Grant}, David and {Greenfield}, Perry and {Groener}, Austen M. and {Guest}, Steve and {Gurovich}, Sebastian and {Handberg}, Rasmus and {Hart}, Akeem and {Hatfield-Dodds}, Zac and {Homeier}, Derek and {Hosseinzadeh}, Griffin and {Jenness}, Tim and {Jones}, Craig K. and {Joseph}, Prajwel and {Kalmbach}, J. Bryce and {Karamehmetoglu}, Emir and {Ka{l}uszy{'n}ski}, Miko{l}aj and {Kelley}, Michael S.~P. and {Kern}, Nicholas and {Kerzendorf}, Wolfgang E. and {Koch}, Eric W. and {Kulumani}, Shankar and {Lee}, Antony and {Ly}, Chun and {Ma}, Zhiyuan and {MacBride}, Conor and {Maljaars}, Jakob M. and {Muna}, Demitri and {Murphy}, N.~A. and {Norman}, Henrik and {O'Steen}, Richard and {Oman}, Kyle A. and {Pacifici}, Camilla and {Pascual}, Sergio and {Pascual-Granado}, J. and {Patil}, Rohit R. and {Perren}, Gabriel I. and {Pickering}, Timothy E. and {Rastogi}, Tanuj and {Roulston}, Benjamin R. and {Ryan}, Daniel F. and {Rykoff}, Eli S. and {Sabater}, Jose and {Sakurikar}, Parikshit and {Salgado}, Jes{'u}s and {Sanghi}, Aniket and {Saunders}, Nicholas and {Savchenko}, Volodymyr and {Schwardt}, Ludwig and {Seifert-Eckert}, Michael and {Shih}, Albert Y. and {Jain}, Anany Shrey and {Shukla}, Gyanendra and {Sick}, Jonathan and {Simpson}, Chris and {Singanamalla}, Sudheesh and {Singer}, Leo P. and {Singhal}, Jaladh and {Sinha}, Manodeep and {Sip{H{o}}cz}, Brigitta M. and {Spitler}, Lee R. and {Stansby}, David and {Streicher}, Ole and {{{S}}umak}, Jani and {Swinbank}, John D. and {Taranu}, Dan S. and {Tewary}, Nikita and {Tremblay}, Grant R. and {Val-Borro}, Miguel de and {Van Kooten}, Samuel J. and {Vasovi{'c}}, Zlatan and {Verma}, Shresth and {de Miranda Cardoso}, Jos{'e} Vin{'i}cius and {Williams}, Peter K.~G. and {Wilson}, Tom J. and {Winkel}, Benjamin and {Wood-Vasey}, W.~M. and {Xue}, Rui and {Yoachim}, Peter and {Zhang}, Chen and {Zonca}, Andrea and {Astropy Project Contributors}},
                title = "{The Astropy Project: Sustaining and Growing a Community-oriented Open-source Project and the Latest Major Release (v5.0) of the Core Package}",
              journal = {\apj},
                 year = 2022,
                month = aug,
               volume = {935},
               number = {2},
                  eid = {167},
                pages = {167},
                  doi = {10.3847/1538-4357/ac7c74},
        archivePrefix = {arXiv},
               eprint = {2206.14220},
         primaryClass = {astro-ph.IM},
               adsurl = {https://ui.adsabs.harvard.edu/abs/2022ApJ...935..167A}
        }

@article{scheel2023multicomponent,
  title={Multicomponent imaging of the Fermi gamma-ray sky in the spatio-spectral domain},
  author={Scheel-Platz, LI and Knollm{\"u}ller, Jakob and Arras, Philipp and Frank, Philipp and Reinecke, Martin and J{\"u}stel, Dominik and En{\ss}lin, Torsten A},
  journal={Astronomy \& Astrophysics},
  volume={680},
  pages={A2},
  year={2023},
  publisher={EDP Sciences}
}

@ARTICLE{sarkar2019possible,
  author  = {{Sarkar}, Kartick C.},
  title   = "{Possible Connection between the Asymmetry of the North Polar Spur and Loop I and Fermi Bubbles}",
  journal = {\mnras},
  year    = 2019,
  month   = feb,
  volume  = {482},
  number  = {4},
  pages   = {4813--4823},
  doi     = {10.1093/mnras/sty2944},
  eprint  = {1804.05634},
  archivePrefix = {arXiv},
  primaryClass  = {astro-ph.GA}
}
\bibliographystyle{aasjournalv7}


\end{document}